\documentclass[conference]{IEEEtran}
\IEEEoverridecommandlockouts
\usepackage{cite}
\usepackage{amsmath,amssymb,amsfonts}
\usepackage{algorithmic}
\usepackage{graphicx}
\usepackage{textcomp}
\usepackage{xcolor}
\usepackage{lipsum}
\usepackage{tabularx}
\usepackage[caption=false]{subfig}
\def\BibTeX{{\rm B\kern-.05em{\sc i\kern-.025em b}\kern-.08em
    T\kern-.1667em\lower.7ex\hbox{E}\kern-.125emX}}
\usepackage{array}
\usepackage{multirow}
\newcolumntype{C}[1]{>{\centering\arraybackslash}p{#1}}
\begin{document}

\begin{figure*}[!t]
\vspace{-18cm}
\noindent © 2024 IEEE. Personal use of this material is permitted. Permission from IEEE must be obtained for all other uses, in any current or future media, including reprinting/republishing this material for advertising or promotional purposes, creating new collective works, for resale or redistribution to servers or lists, or reuse of any copyrighted component of this work in other works.
\end{figure*}

\newpage
\title{Using Capability Maps Tailored to Arm Range of Motion in VR Exergames for Rehabilitation 
\thanks{The authors are with the Mechanical and Aerospace Engineering Department, NYU Tandon School of Engineering, Brooklyn, NY 11201, USA (corresponding author: 646-997-3161; e-mail: vkapila@nyu.edu).}
}

\author{Christian Lourido, Zaid Waghoo, Hassam Khan Wazir, Nishtha Bhagat, and Vikram Kapila}

\maketitle

\begin{abstract}
Many neurological conditions, e.g., a stroke, can cause patients to experience upper limb (UL) motor impairments that hinder their daily activities. 
For such patients, while rehabilitation therapy is key for regaining autonomy and restoring mobility, its long-term nature entails ongoing time commitment and it is often not sufficiently engaging.  
Virtual reality (VR) can transform rehabilitation therapy into engaging game-like tasks that can be tailored to patient-specific activities, set goals, and provide rehabilitation assessment.
Yet, most VR systems lack built-in methods to track progress over time and alter rehabilitation programs accordingly.
We propose using arm kinematic modeling and capability maps to allow a VR system to understand a user's physical capability and limitation.
Next, we suggest two use cases for the VR system to utilize the user's capability map for tailoring rehabilitation programs.
Finally, for one use case, it is shown that the VR system can emphasize and assess the use of specific UL joints. 

\indent \textit{Clinical relevance}—This paper's VR-based system can tailor a rehabilitation tool to a user's capability and limit. 


\end{abstract}

\vspace{-0.4em}
\section{Introduction}\label{sec:introduction}
\vspace{-0.4em}
Stroke is the second most common cause of death worldwide \cite{feigin_global_2017}, and survivors are likely to suffer from upper limb (UL) motor impairments \cite{parker_loss_1986} that prevent them from effectively performing activities of daily living (ADLs) \cite{sze_chit_leong_examining_2022}.
Long-term rehabilitation therapy is essential in helping patients restore the mobility of the UL and regain autonomy to perform ADLs. 
However, the rehabilitation process can be time-consuming and not engaging, resulting in non-compliance by patients in performing the exercises prescribed for specified periods \cite{viglialoro_review_2019}.

Virtual reality (VR) based systems allow the design of rehabilitation activities as engaging tasks in the form of computer games that effectively keep patients motivated during the rehabilitation process \cite{budziszewski_low_2013}. 
Using VR-based rehabilitation with appropriate assessment tools for UL allows therapists to tailor patient-specific activities, set performance goals, and monitor rehabilitation progression \cite{kim_assessment_2022}.
Due to the long-term nature of rehabilitation, access to a therapist is critical for effectively assessing patient progress. 
Data-driven VR-based exergames can aid therapists in providing a holistic assessment while optimizing the rehabilitation assessment process. 

With recent advancements in human pose estimation techniques, several solutions for UL range of motion (ROM) assessment have been proposed, e.g., the use of wearable sensors \cite{rajkumar_wearable_2020}, optical solutions \cite{wazir_range_2022}, and virtual, augmented, and mixed reality (VAMR) systems \cite{sveistrup_motor_2004}. 
The latter primarily use head-mounted displays (HMDs) and are supplemented with built-in inertial measurement units (IMUs) or external optoelectronic systems, allowing accurate measurement of a user's pose.
Endowing VAMR systems with ROM estimation features enables them to estimate relevant kinematic variables (e.g., joint angles), which can identify UL deficits, quantify impairment levels, and complement a clinical assessment \cite{subramanian_validity_2010}.

Researchers \cite{xie_innovative_2021,zanchettin_kinematic_2011, bertomeu-motos_human_2018} have also employed analytical methods, such as the kinematic modeling of the human arm, to objectively comprehend the ROM of a user's UL, particularly for impaired limb motion.
The mathematical modeling using a kinematic chain, informed by the degrees of freedom (DoFs) and limb dimensions, describes the workspace, positional capabilities, and constraints of the user's UL \cite{jerbic_35_2020}.
This method permits estimating the user's workspace by considering the joint ROM limits due to a neurological condition. 
Alternatively, experimentally evaluating and analyzing the achievable workspace can aid in assessing the user's health \cite{hernandez_novel_2021}  and formulating effective therapy interventions tailored to the user.

We propose a system that combines a user's experimentally computed workspace with a VR rehabilitation exergame to provide 
exercises tailored to the capabilities and limitations of the user's UL. 
We restrict the user's right shoulder with a shoulder brace to simulate a shoulder restriction and measure their resulting ROM. 
A Kinect sensor collects the limb dimensions and ROM of UL joints as the user  performs a series of UL exercises to help create a capability map \cite{zacharias_capability_map_2007} for the simulated disability condition.
Next, we use the information from the capability maps to assess the user's condition and tailor activities in the VR rehabilitation exergame.

The paper is organized as follows. Section \ref{sec:methodology} describes the methods employed in the design of the proposed rehabilitation system framework. 
Section \ref{sec:sysevalanddiscussion} presents the results of the rehabilitation system tests. 
Section \ref{sec:conclusion} draws some conclusions on the findings and suggests directions for future research.

\vspace{-1.0em} 
\section{METHODOLOGY} \label{sec:methodology}
\vspace{-0.8em} 
Four users are recruited to voluntarily test the VR system. 
Each user with an UL restriction performs a set of configuration exercises (see Section \ref{sec:sysevalanddiscussion} for details). 
The resulting measurements provide limb lengths and restrict ROM of UL joints used with a human arm kinematic model to determine the user's workspace. 
Then, using the capability analysis \cite{zacharias_capability_map_2007}, we determine regions where each user would be more dexterous. 
This information is utilized to adapt the tasks in the VR exergame for a particular user.

\vspace{-0.8em} 
\subsection{Human Arm Kinematic Model}
\begin{figure}[!t]
  \centering
  \subfloat[]{
  \includegraphics[height=3.5cm]{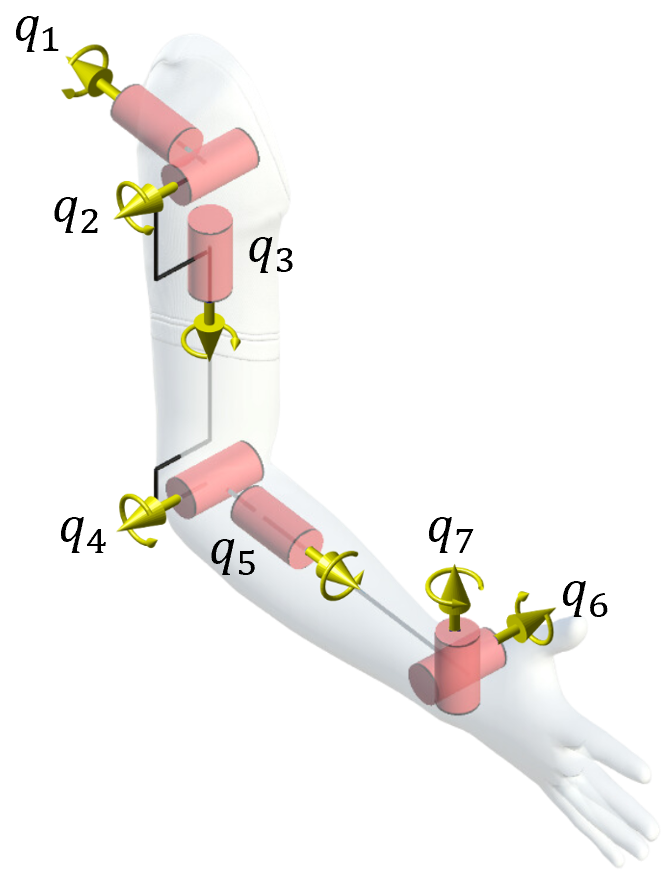}
  \label{fig:arm-kinematic-model}
  }
  \subfloat[]{
  \includegraphics[height=3.5cm]{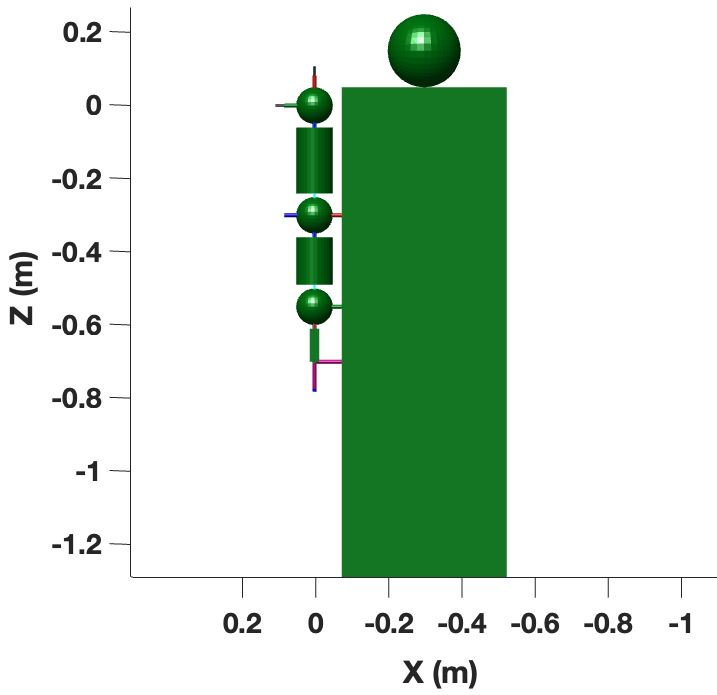} 
  \label{fig:rigid-body-tree-meshes}}
  \vspace{-0.3em}  
  \caption{(a) Arm kinematic model and (b) rigid-body tree with collision meshes.}
  \label{fig:models}
  \vspace{-1.8em} 
\end{figure}

\vspace{-0.4em} 
The human arm can be simplified into a 7-DoF kinematic model (\hspace{-0.01em}\cite{zanchettin_kinematic_2011},\hspace{-0.01em}\cite{bertomeu-motos_human_2018}). 
In this model, the glenohumeral joint in the shoulder complex is characterized as a spherical joint and modeled as three orthogonal revolute joints corresponding to shoulder abduction-adduction ($q_1$), flexion-extension ($q_2$), and internal-external rotation ($q_3$).
Similarly, the elbow joint is modeled as two orthogonal revolute joints
corresponding to elbow flexion-extension ($q_4$) and forearm pronation-supination ($q_5$). Note that $q_5$ is considered a wrist DoF  \cite{bertomeu-motos_human_2018}. 
Finally, two additional DoFs are modeled in the wrist: the ulnar-radial deviation ($q_6$) and flexion-extension ($q_7$). Fig. \ref{fig:arm-kinematic-model} shows the model used in this work.

\vspace{-0.8em}
\subsection{Capability Map}
\vspace{-0.3em}

To assess and understand the capabilities of the arm kinematic model, we analyze the arm's reachable workspace, which comprises the set of poses in $\mathbb{R}^6$ (position and orientation) that the kinematic structure can reach, given some joint angle values \cite{porges_reachability_2014}.
To do so, the hybrid method of \cite{porges_reachability_2014} is followed, which discretizes the workspace, describes it as a voxelized structure, and uses forward and inverse kinematics to assign a binary value to indicate if a voxel is reachable or not, thus creating a reachability map \cite{diankov_reachability_2010}. 
Finally, a capability map is built by assigning a reachability score to each voxel based on the number of different orientations reachable given a position of the tool center point (TCP) of the kinematic structure \cite{zacharias_capability_map_2007}. 
For the arm model of Fig. \ref{fig:arm-kinematic-model}, the TCP is defined at the tip of the user's index finger.
We use MATLAB to generate the capability maps by modeling the arm kinematics using a rigid-body tree structure and incorporating links representing the upper arm and forearm (see Fig. \ref{fig:rigid-body-tree-meshes}). 
Additionally, the model is augmented with volumetric collision meshes, allowing the system to consider both arm-arm and arm-body collisions during the map generation process and discard the regions from the capability map where a collision is detected. 

The system is configured with the user's upper limb link lengths that are determined using the distances between upper limb joints. 
These data and the ROM of each joint provide the system with the necessary information to generate the capability map.
Note that the generation of the capability map is performed offline, and its data is queried online when the VR exergame system runs. 
This is necessary since generating the map requires processing a large amount of data. 

\begin{figure}[!t]
  \centering
  \subfloat[]{
  \includegraphics[height=3.6cm]{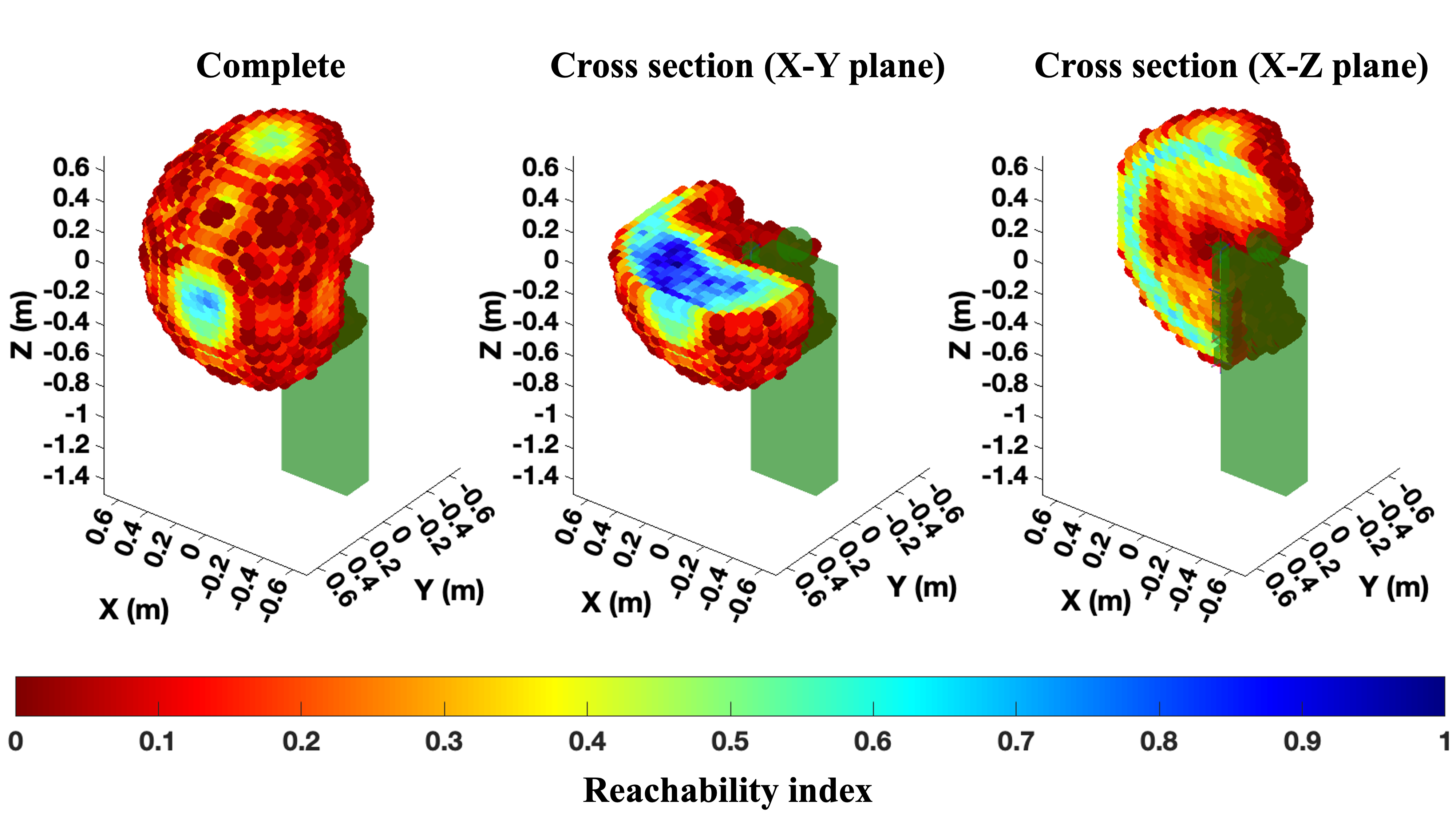}
  \label{fig:cm-healthy}
  }
  	\vspace{-0.2em} 
  \subfloat[]{
  \includegraphics[height=3.6cm]{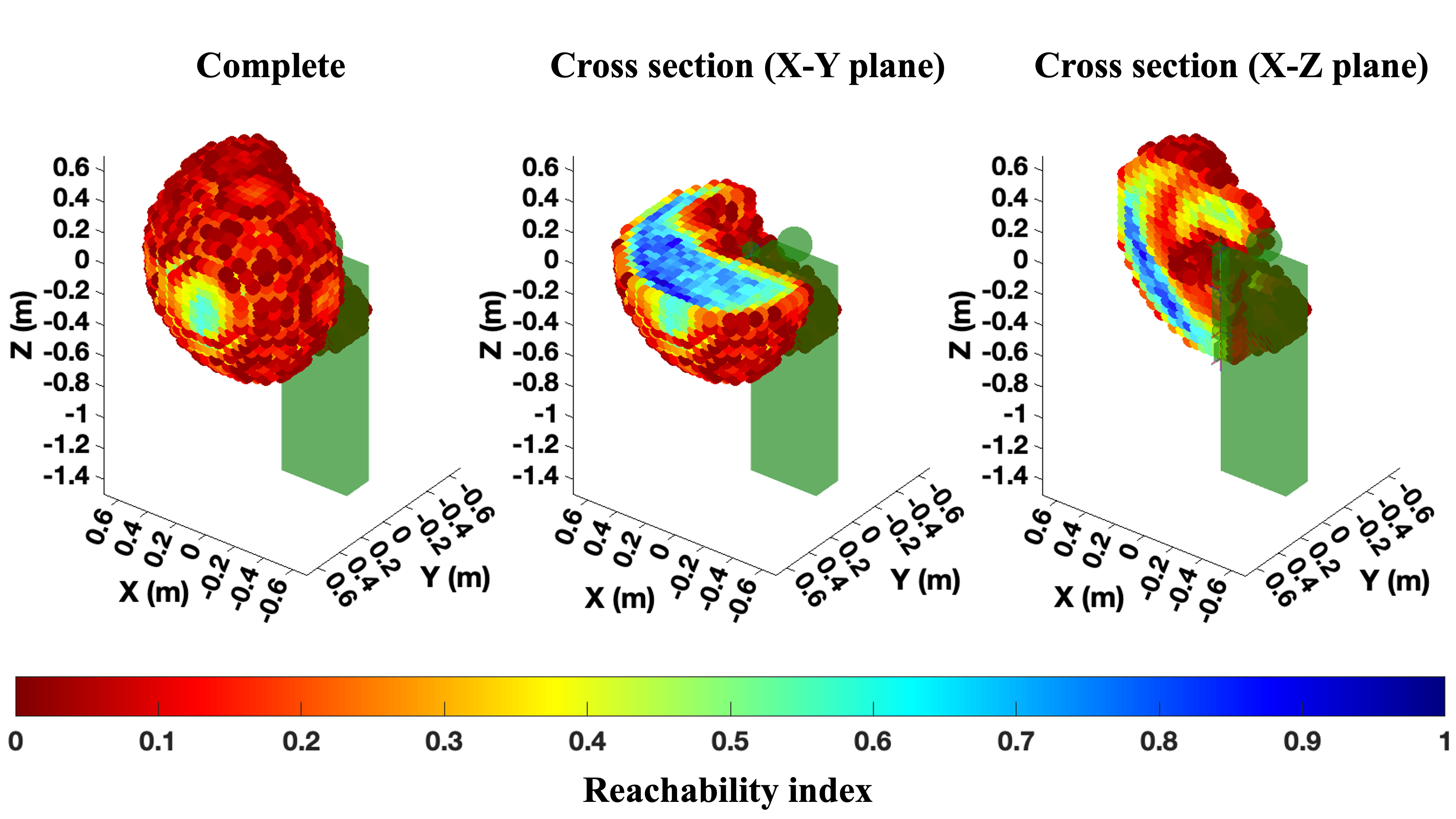}
  \label{fig:cm-restricted}}
\vspace{-0.5em}  
  \caption{Capability maps: (a) healthy user and (b) user wearing a restriction.}
  \label{fig:capability-maps}
  \vspace{-2em}
\end{figure}

The VR exergame system requires two capability maps.
The first map uses the nominal values for joint ROM available from the Centers for Disease Control and Prevention (CDC) \cite{soucie_range_2011} and related research (\hspace{-0.01em}\cite{rajkumar2021},\hspace{-0.01em}\cite{kim_wrist_2014}), and it represents information about the dexterity of a healthy user in their reachable workspace (see Fig. \ref{fig:cm-healthy}). 
\color{black} The calculated ROM of the joints during the configuration exercises by users with UL restriction helps create the second map, which contains information about the dexterity of the user's UL in their reachable workspace under their current condition. 
For this work, we simulate the restricted shoulder condition using a shoulder brace (see Fig. \ref{fig:cm-restricted}). 
The information from both maps is used to configure the parameters of the VR exergame system for each user.

\vspace{-1em}
\subsection{Rehabilitation Exergame}
\vspace{-0.5em}
The rehabilitation exergame has three key components: (i) ROM measurement, (ii) a VR exergame for ROM improvement, and (iii) a VR exergame for occupational therapy (OT).

\subsubsection{ROM measurement}
Typically, a patient's ROM is determined by a physical therapist through a comprehensive ROM assessment. 
The therapist guides the individual through specific exercises and uses a goniometer to measure the achieved ROM for each assessed joint \cite{soucie_range_2011}. 
The current work uses the Kinect sensor to automate the ROM measurement.
The user stands in a {\it{neutral}} pose in front of the Kinect with arms on their side for the system to detect their UL joints in {3D}.
From the {\it{neutral}} pose data, the system calculates the limb lengths using the distances between the joints detected by the Kinect.
Next, the user with UL restriction performs UL-ROM exercises, and the maximum and minimum achieved angles for each movement are recorded.
The data of limb lengths and ROM is fed to the human arm kinematic model to generate the capability maps for the user. 
The nominal ROM data 
(\hspace{-0.3em}~\cite{soucie_range_2011},\hspace{-0.01em}\cite{rajkumar2021}) is used to generate the {\it{nominal}} or healthy capability map.
Similarly, the measured ROM data for the user with UL restriction generates the {\it{restricted}} capability map.

\begin{figure}[!t]
    \vspace{-1em} 
  \centering
  \subfloat[]{
    \raisebox{-0.5\height}{
      \includegraphics[height=2.8cm]{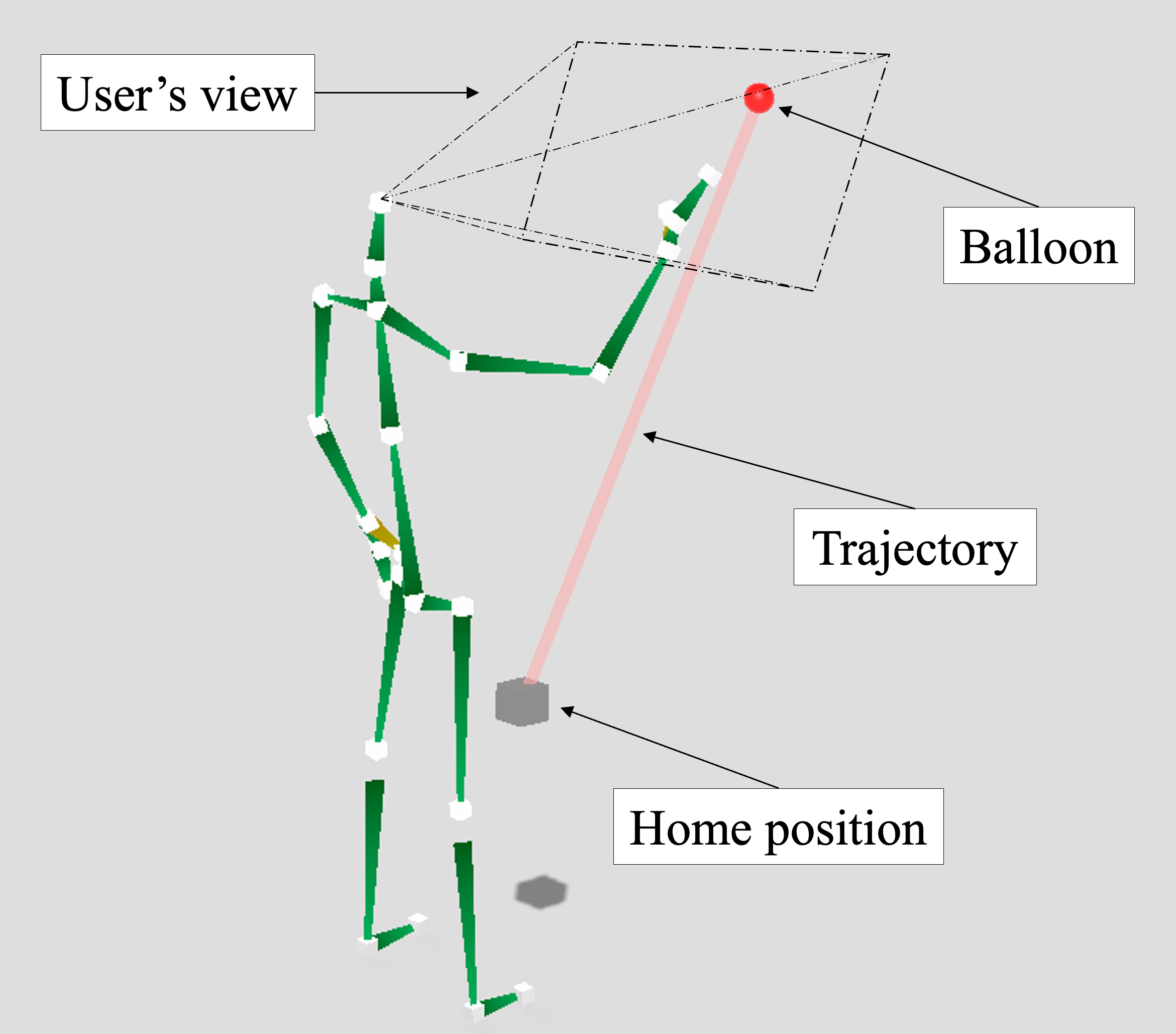}
    }
    \label{fig:rom-game-therapist}
  }
  \subfloat[]{
    \raisebox{-0.5\height}{
      \includegraphics[height=2.8cm]{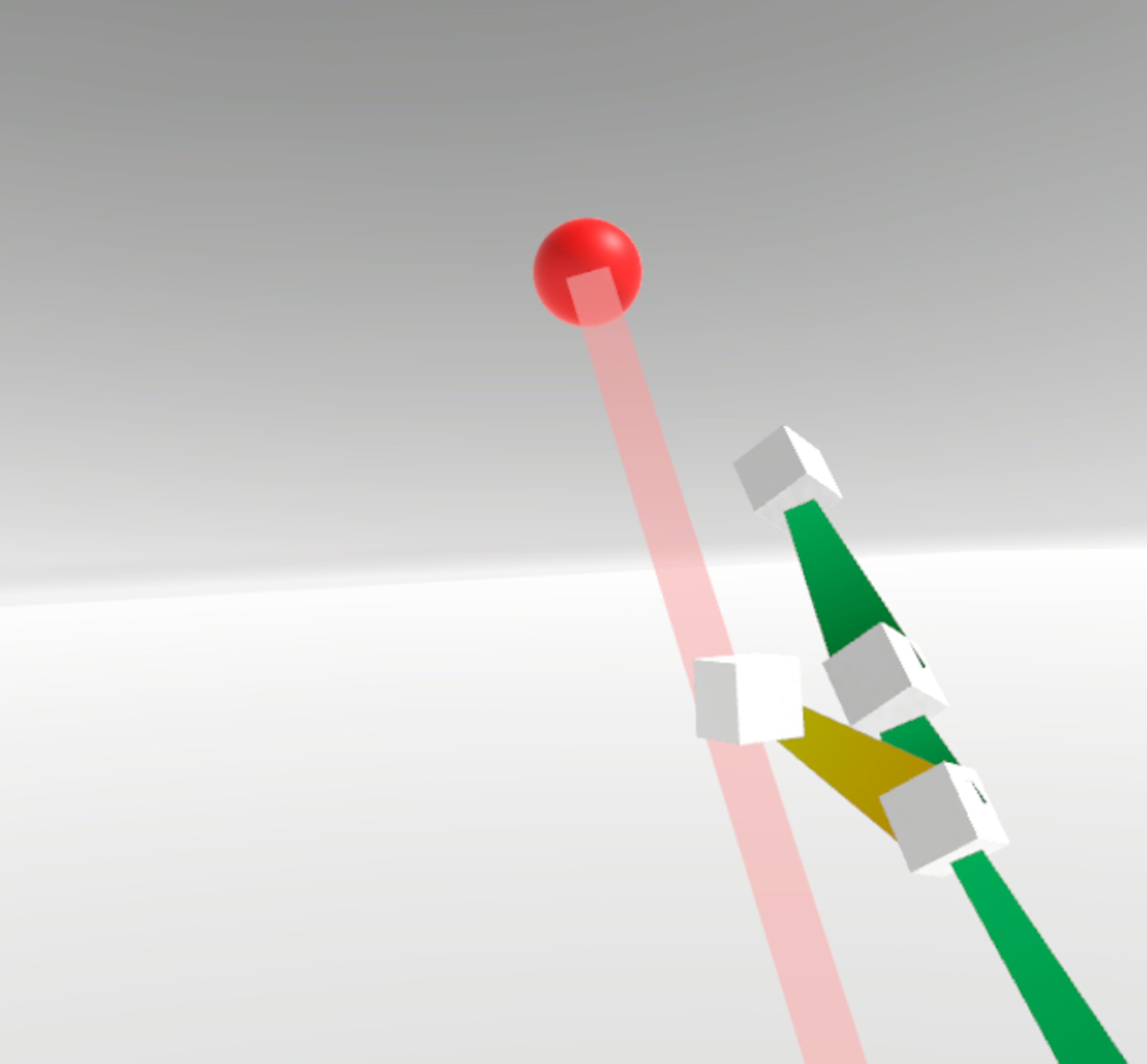}
    }
    \label{fig:rom-game-user}
  }
    \vspace{-0.5em}  
  \caption{Exergame for ROM improvement (a) therapist view and (b) user view.}
  \label{fig:Exergame-ROM-improvement}
    \vspace{-1.8em}
\end{figure}

\subsubsection{VR exergame for ROM improvement}
This VR exergame is proposed to aid the design of ROM exercise programs.
It requires the user to wear a VR headset and stand at a distance from the Kinect, where their body joints can be tracked. 
The therapist starts the game and loads the user's previously generated capability map into the game.
Next, the user is presented with a virtual balloon floating in  pre-defined regions within their capability map. 
The therapist is expected to define these regions by using the reachability information as one of the criteria.
The user is tasked with popping the balloon by reaching out and touching it with their hand by following a virtual trajectory from a {\it{home}} position to the balloon.  
Based on a repetition number set by the therapist, balloons are spawned, one at a time, at various reachable locations. 
The balloon location and virtual trajectory can be selected to focus on operationalizing specific joints of the UL. 
The goal is to replicate, in an engaging manner, the repetitive movements inherent in physical therapy, mirroring the therapeutic efficacy of such motions \cite{tseng2007Comm}. 
This intentional approach seeks to enhance the user's ROM by simulating and encouraging the repetition of specific movements within the virtual environment tailored to the user's capability. 
Fig. \ref{fig:Exergame-ROM-improvement} shows the images of the proposed exergame, with views from the perspectives of the therapist and the user.

\subsubsection{VR exergame for OT}
\begin{figure}[!t]
    \vspace{-1em} 
  \centering
  \subfloat[]{
    \raisebox{-0.5\height}{
      \includegraphics[height=2.9cm]{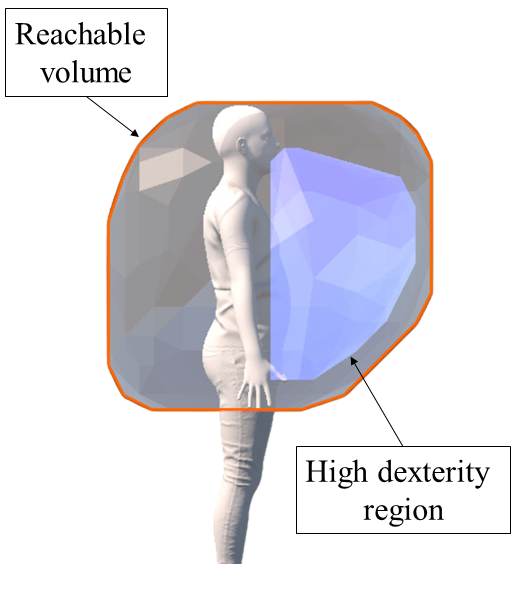}
    }
    \label{fig:visual-cues}
  }
  \subfloat[]{
    \raisebox{-0.5\height}{
      \includegraphics[height=2.7cm]{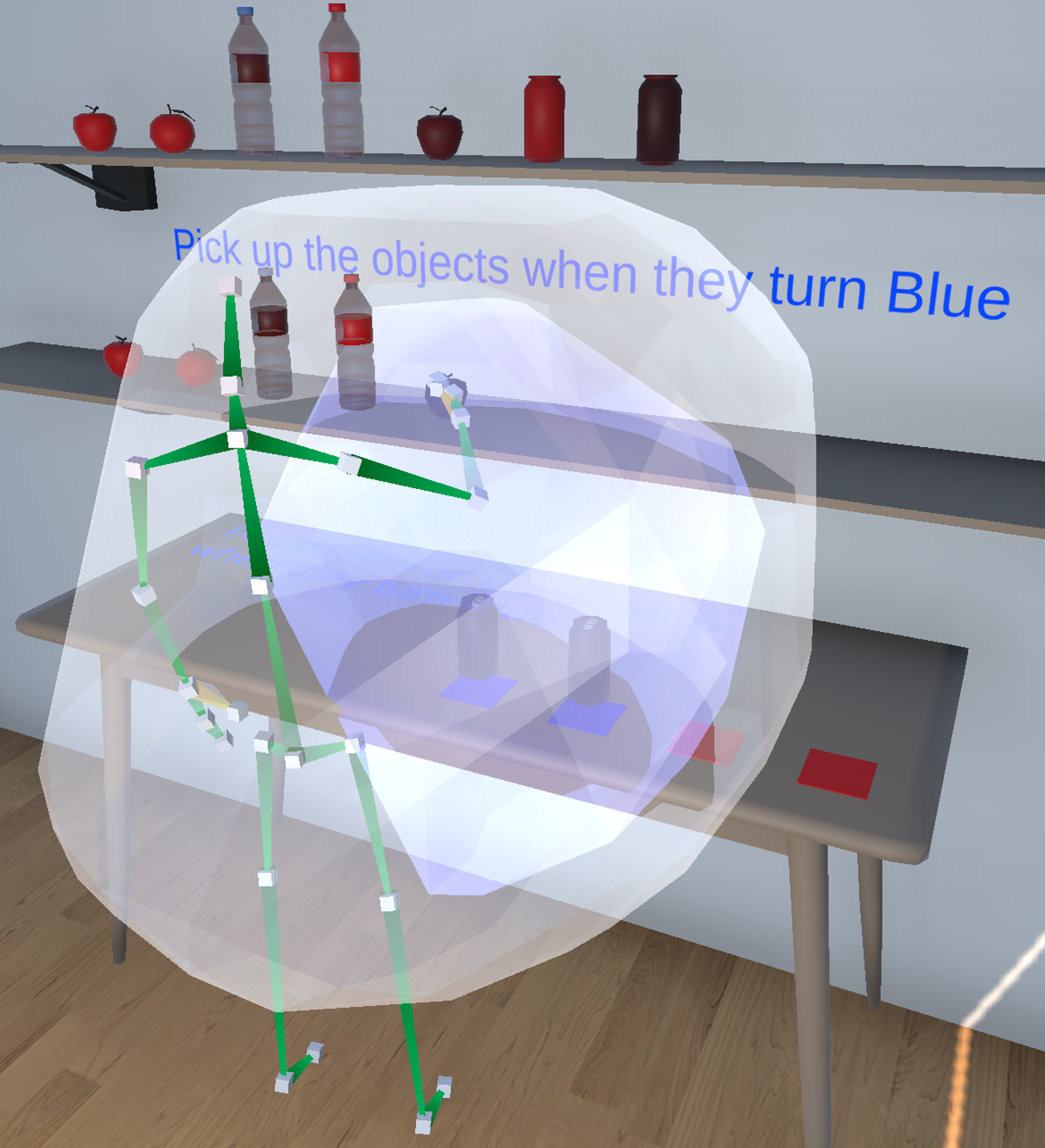}
    }
    \label{fig:otgame-therapist}
  }
  \subfloat[]{
    \raisebox{-0.5\height}{
      \includegraphics[height=2.7cm, trim={3cm 0 2cm 0},clip]{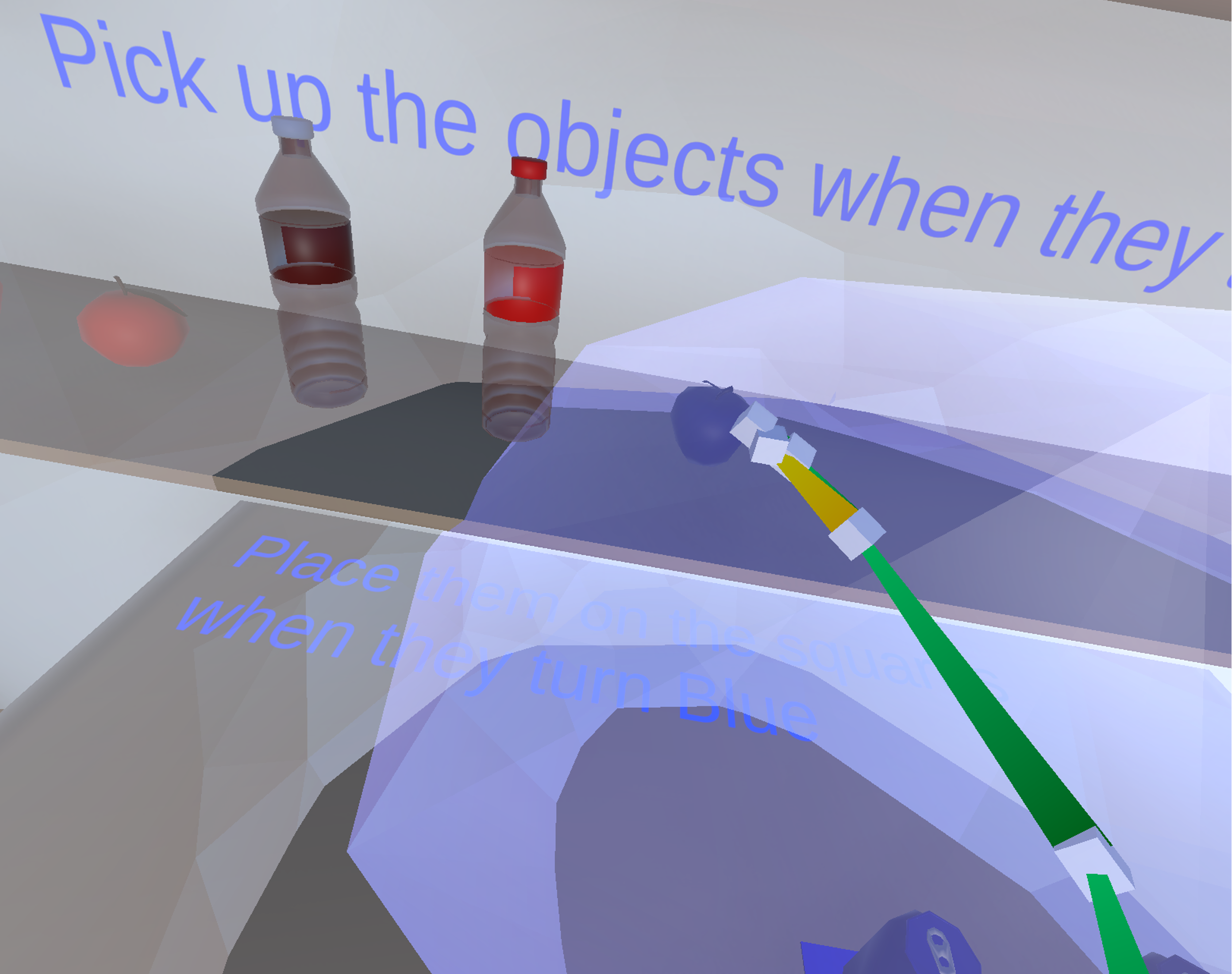}
    }
    \label{fig:otgame-user}
  }
    \vspace{-0.5em}  
  \caption{Exergame for OT (a) visual cues, (b) therapist view, and (c) user view.}
  \label{fig:occupationaltherapygame}
    \vspace{-1.8em}
\end{figure}

In addition to ROM improvement, the capability maps can be used for OT interventions. 
This is illustrated using a VR exergame that harnesses individualized capability maps to craft visual cues for users performing specific OT tasks. 
The exergame provides users with a comprehensive view of their reachable workspace in the virtual environment. 
The VR system uses the information from the \textit{restricted} capability map and a convex hull algorithm \cite{Unity_quickhull2018} to find polyhedrons that enclose voxels within a specified reachability score range.
For this exergame, one visual cue is created using a polyhedron enclosing all the voxels of the restricted capability map to show the user's complete reachable workspace. 
Another visual cue comprises a polyhedron surrounding the voxels with the highest reachability score, corresponding to a range defined by the therapist who designs the rehabilitation program (see Fig. \ref{fig:visual-cues}).
These cues make users aware of their reachable workspace and the regions where they would be more dexterous and help them craft strategies to grasp virtual objects during the exergame. 
The exergame also uses color changes to accentuate objects within the user's workspace, thus signaling to the user that they are within reach for grasping. 
As the user moves around in the environment, the VR system emphasizes the virtual objects as they appear situated in regions of higher reachability, which indicates the optimal zones for grasping. 
The exergame tasks the user to pick up virtual objects from a virtual shelf and place them on a virtual table within designated spots.
This scenario mirrors ADLs, such as gathering ingredients to prepare a meal or arranging dishes, thus creating a realistic and practical OT environment tailored to the user's capability.
Figs. \ref{fig:otgame-therapist} and \ref{fig:otgame-user} show the exergame for OT, as seen by the therapist and the user, respectively. 

\vspace{-0.5em}
\section{SYSTEM EVALUATION} \label{sec:sysevalanddiscussion}
\vspace{-0.55em}
 
To evaluate the proposed system and analyze the impact of simulated shoulder-restriction conditions, we gathered data from four users who tested the exergame tailored to them for ROM improvement. We assessed each user's mobility, specifically regarding their arm-reachable workspace, dexterity, and speed. 
To simulate varying degrees of shoulder mobility, we used a shoulder brace to create three scenarios: unrestricted, restricted, and partially restricted shoulder movements. 
The unrestricted shoulder case represents a user with a healthy condition, the restricted shoulder case simulates a user with a shoulder injury, and the partially restricted case simulates a user experiencing mobility recovery due to rehabilitation. 

Each user stood in front of the Kinect in the restricted and partially restricted conditions and performed the shoulder (abduction-adduction, flexion-extension, internal-external rotation) and elbow (flexion-extension) exercises. 
The system measured the ROM for joint movements $q_i$, $i = 1, \ldots, 4$. For $q_i$, $i = 5, \ldots, 7$, the system was provided the ROM values used for creating the {\it{nominal}} capability maps since Kinect can not precisely estimate the ROM of wrist joints. 
Alternatively, a therapist can use a goniometer or other methods to accurately measure the ROM of these joints and use it with the VR exergame.
Once the data for each user is collected, the system computes their capability map for each scenario.

\subsubsection{Reachable volume and dexterity analysis}
We assessed the workspace volume reduction due to shoulder restriction by comparing the number of occupied voxels in the capability maps for the healthy {\it{vs}} the partially restricted and restricted shoulder cases. 
Next, we examined the reachability scores of the common voxels in the capability maps of healthy and partially restricted cases, as well as healthy and restricted cases. 
By adding the scores of all voxels in these common regions, the overall reachability scores were obtained, allowing the computation of dexterity reduction.  
Table \ref{table:volume_deterity} lists the reductions in volume and dexterity (in \%) for the partially restricted and  restricted cases {\it{vs}} the unrestricted case for all users.

\begin{table}[t!]
    \vspace{-1.5em}
    \centering
    \caption{Volume and dexterity reduction due to shoulder restriction}
    \vspace{-0.5em}
    \label{table:volume_deterity}
    \begin{tabular}{
    |c| 
    C{1.4cm}| 
    C{1.4cm}| 
    C{1.4cm}| 
    C{1.41cm}| 
    } 
    \hline
    \multirow{2}{*}{\centering}& \multicolumn{2}{c|}{Partially Restricted} & \multicolumn{2}{c|}{Restricted} \\ \cline{2-5}
     \multirow{1}{*}{\centering ID}& \% Volume Reduction & \% Dexterity Reduction & \% Volume Reduction & \% Dexterity Reduction \\ \hline
    1 & 11.17   & 0.78  & 27.80 & 11.50 \\ 
    2 & 11.11   & 2.54  & 21.20 & 5.34\\ 
    3 & 7.80    & 2.88  & 29.92 & 10.07 \\ 
    4 & 10.00   & 5.17 & 28.94 & 15.65 \\ \hline 
    \end{tabular}
    \label{table:volume_deterity}
\end{table}

\subsubsection{Tailored ROM exergame}
Using the data from the capability maps, we designed a ROM exergame activity involving popping balloons in distinct regions. The regions were classified into three levels of difficulty: easy, medium, and hard, based on the reachability score of voxels from those regions. After wearing the VR headset and standing before the Kinect, the user started the exergame.  
With an egocentric perspective, the user identified their body as a stick figure in the VR scene and moved their right wrist to the home position, visualized in the game as a virtual cube.
From this home position, the user moved their wrist to the location of the balloon displayed in the VR scene to pop it. 
To assess shoulder mobility, the user was instructed to follow a virtual straight-path trajectory while keeping their right arm as straight as possible. This aspect of the exergame design emphasized the use of the shoulder joint. During the activity, the VR system used the distance between the home position and the displayed balloon and the time taken to pop the balloon to estimate the speed of each pop.

\subsubsection{Results of the tailored ROM exergame}
The users were tasked to pop 10 balloons in each difficulty level and under each shoulder condition: restricted, partially restricted, and unrestricted, sequentially. 
With one exception, results in Table \ref{table:speeds} show a trend of increase in balloon-popping speed with the lessening of difficulty and shoulder restriction levels. 

\begin{table}[t!]
    \vspace{-1.5em}
    \centering
    \caption{Reachable volume and dexterity analysis}
    \vspace{-0.5em}
    \label{table:speeds}
    \begin{tabular}{
    |C{0.2cm}| 
    C{0.46cm}| 
    C{0.46cm}| 
    C{0.46cm}| 
    C{0.46cm}| 
    C{0.46cm}| 
    C{0.46cm}| 
    C{0.46cm}| 
    C{0.46cm}| 
    C{0.46cm}| 
    } 
    \hline
    \multirow{2}{*}{\centering}& \multicolumn{3}{c|}{Unrestricted} & \multicolumn{3}{c|}{Partially Restricted} & \multicolumn{3}{c|}{Restricted} \\ \cline{2-10}
    ID & Easy (m/s) & Med. (m/s) & Hard (m/s)& Easy (m/s) & Med. (m/s) & Hard (m/s) & Easy (m/s) & Med. (m/s) & Hard (m/s) \\ \hline
    1 & 0.66 & 0.50 & 0.44 & 0.51 & 0.43 & 0.40 & 0.32 & 0.28 & 0.26\\ 
    2 & 0.69 & 0.41 & 0.37 & 0.45 & 0.38 & 0.33 & 0.28 & 0.26 & 0.29\\ 
    3 & 0.34 & 0.24 & 0.23 & 0.26 & 0.24 & 0.22 & 0.24 & 0.21 & 0.18\\ 
    4 & 0.57 & 0.41 & 0.32 & 0.37 & 0.32 & 0.29 & 0.26 & 0.22 & 0.19\\  \hline  
    \end{tabular}
    \label{table:speeds}
    \vspace{-1.5em}
\end{table}

\vspace{-0.6em}
\section{CONCLUSION AND FUTURE WORK} \label{sec:conclusion}
\vspace{-0.6em}

This paper introduced a rehabilitation method that uses arm kinematic model and capability maps to enable a VR exergame system to understand a patient's physical capability and limitation.
Under this framework, the exergames for ROM improvement and OT can be tailored for a user while objectively assessing their performance. 
Our preliminary evaluation captured the ROM data of four users in varied simulated scenarios to demonstrate the system's utility by analyzing metrics related to the users' reachable workspace and dexterity. Moreover, for the ROM exergame, we demonstrated the ability to emphasize and assess the movement capability of a specific joint of UL. Future research will explore the user experience aspect of the VR system and expand the set of metrics derived from capability maps. 
These metrics may be crucial in effectively communicating a user's condition for enhancing the VR-based rehabilitation's overall impact.

\bibliographystyle{IEEEtran}

\vspace{-0.9em}
\bibliography{references}
\end{document}